\documentclass[conference]{IEEEtran} 
\usepackage{amsmath,amssymb,amsfonts}
\usepackage{cite}
\usepackage{bm}
\usepackage{mathtools}
\usepackage[binary-units=true]{siunitx}
\DeclareSIUnit{\belmilliwatt}{Bm}
\usepackage{graphicx}
\usepackage{dsfont}
\usepackage{subcaption}
\usepackage{comment}
\usepackage{algorithm}
\usepackage{algpseudocode}
\usepackage{mathtools}

\usepackage{textcomp}
\usepackage[export]{adjustbox}
\usepackage{float}
\usepackage{booktabs}
\usepackage{multicol}
\usepackage{xparse}
\usepackage{color,soul}
\usepackage[bottom]{footmisc}
\usepackage[binary-units=true]{siunitx}
 \usepackage{bm}           
\usepackage{tikz}         
\usetikzlibrary{arrows.meta, positioning, calc}
\DeclareSIUnit{\belmilliwatt}{Bm}
\DeclareSIUnit{\dBm}{\deci\belmilliwatt}
\DeclareSIUnit[per-mode=symbol,per-symbol=p]{\Bps}{\byte\per\second}

\sethlcolor{yellow}
\usepackage{algpseudocode}

\usepackage{hyperref}

\makeatletter
\def\BState{\State\hskip-\ALG@thistlm}
\makeatother

\IEEEoverridecommandlockouts
\begin{document}

    \title{Perception-Aware Joint Power and Sub-Band Allocation for 6G In-Body Subnetworks}

\author{
\IEEEauthorblockN{Samira Abdelrahman\IEEEauthorrefmark{1} and Hossam Farag\IEEEauthorrefmark{2}\IEEEauthorrefmark{1}}
\IEEEauthorblockA{
\IEEEauthorrefmark{1} Department of Electrical Engineering, Aswan University, Egypt\\
\IEEEauthorrefmark{2}Department of Electronic Systems, Aalborg University, Denmark \\
Email: sma@asw.edu.eg,  hmf@es.aau.dk
}
}

	\maketitle
	
	\begin{abstract}  
In-body subnetworks (IBSs) are expected to become a key enabler of immersive eXtended Reality (XR) services in sixth-generation (6G) networks by providing ultra-short-range, low-latency wireless connectivity around the human body. However, the dense coexistence of multiple IBSs leads to severe co-channel interference, requiring increased transmit power to satisfy the stringent latency requirements of XR applications. Existing interference management approaches allocate radio resources solely according to application-level Quality-of-Service (QoS) requirements, overlooking the perceptual limitations of human users. This paper proposes perception-aware joint power control and sub-band allocation framework that integrates users' delay perception into radio resource allocation for XR-oriented IBSs. A learning-based perception model is first developed by combining Gaussian mixture modeling (GMM) with supervised learning to develop a statistical model of the delay perception threshold. The learned perception model is then incorporated into a stochastic radio resource allocation problem, which is reformulated using a Lyapunov drift-plus-penalty and solved through a low-complexity per-slot resource allocation procedure. System-level simulations under realistic intra- and inter-IBS propagation conditions demonstrate that the proposed approach substantially improves radio resource efficiency, achieving up to 26\% transmit power reduction under stringent latency requirements and approximately 60\% power savings in dense IBS deployments, while maintaining the required Quality of Experience (QoE).

	\end{abstract}
\section{Introduction}\label{sec:intro}
The sixth generation (6G) of wireless networks is envisioned to support a new class of extreme connectivity scenarios through the concept of in-X subnetworks~\cite{IN-X1}: low-power, short-range, wireless cells embedded within or in close proximity to physical entities such as the human body (in-body), industrial machinery (in-machine), or vehicular platforms (in-vehicle). Of particular relevance among in-X subnetwork deployments are in-body subnetworks (IBSs), which enable short-range, localized wireless connectivity in close proximity to the human body~\cite{XR}. A compelling and representative use case for IBSs is eXtended Reality (XR), which is gaining high potential in various domains including health~\cite{rehab} (guided rehabilitation), education~\cite{educ} (immersive learning) and entertainment~\cite{avatar} (shared XR events). By confining traffic exchanges to short ranges and leveraging local processing capabilities, IBSs can support dense deployments of XR users. However, as the number of XR users increases, interference becomes the dominant factor~\cite{interference} limiting the reliability of high-rate XR video delivery (poor user experience), forcing conventional resource allocation schemes to increase transmit power in order to satisfy application-level QoS constraints. i.e., lower delay and/or higher data rate. 

Several studies have focused on interference mitigation in the context of industrial/factory 6G subnetworks~\cite{sub1, sub2, sub3, sub4}. In~\cite{sub1}, a centralized radio resource management framework is introduced, where sub-bands are assigned through an iterative sequential optimization procedure that aims to minimize the aggregate interference-to-signal ratio across the network. A deep learning-based resource allocation strategy is presented in~\cite{sub2}, targeting sub-band assignment to maximize the number of subnetworks that satisfy heterogeneous data rate requirements. To address distributed radio resource management, \cite{sub3} develops a multi-agent deep reinforcement learning framework for non-coordinated in-X subnetworks. The proposed solution employs recurrent neural networks to capture temporal variations in wireless channels, while a binary-tree search mechanism is incorporated to improve communication reliability. For autonomous subnetwork operation, the work in~\cite{sub4} presents an adaptive sub-band allocation scheme that seeks to minimize spectrum utilization while reducing the frequency of subnetwork reconfiguration between transmission intervals. Furthermore, \cite{sub5} proposes a decentralized interference management approach that jointly considers application goals and control requirements to coordinate interference among subnetworks operating in industrial environments. 

In the context of IBSs, there are limited works~\cite{IBS1, IBS2, interference, IBS4} focusing primarily on developing adaptive interference management algorithms where IBS users coexist with other cellular users.  However, such approaches neglect the human factor, more specifically,  the perceptual characteristics of human users, which are expected to play a critical role in emerging XR-oriented consumer subnetworks. Previous studies~\cite{human1, human2} have demonstrated that the cognitive limitations of the human brain fundamentally influence how wireless users perceive network performance, resulting in a nonlinear relationship between objective Quality of Service (QoS) metrics and the experienced Quality of Experience (QoE). Specifically, improvements in transmission latency below the user's perceptual threshold do not necessarily translate into a better quality of experience. Consequently, allocating additional transmit power to combat interference beyond what is perceptible to the user results in inefficient spectrum and energy utilization. These observations motivate the development of human-aware interference management, where transmit power is optimized not only according to wireless channel conditions and interference levels, but also according to the cognitive delay perception of individual XR users, thereby achieving reliable XR communication while significantly improving radio resource efficiency. This way, the AP can avoid wasting radio resources on QoS gains that are imperceptible to the human users. 

Motivated by this gap, this paper proposes a human-aware power control framework for  6G IBSs that jointly exploits wireless network dynamics and the cognitive characteristics of end users. Our work is inspired by  the psychophysical concept of a just-noticeable difference, i.e., the minimum change in a stimulus that a human observer is able to detect~\cite{JND}. Applied to network-induced latency, we introduce the just-noticeable delay (JND),
which marks the delay threshold below which further delay reduction yields no discernible improvement in experienced quality, regardless of how much additional radio resource the network expends to achieve it. Crucially, unlike the fixed, application-level latency targets assumed by conventional QoS-driven schemes, the JND is inherently user-specific and time-varying, shaped by factors such as cognitive state, interaction intensity, and prior exposure to the XR content. We formulate a perception-aware radio resource allocation problem that minimizes the long-term transmit power while guaranteeing probabilistic QoE requirements determined by each user's JND rather than fixed application-level latency constraints. We first develop a machine learning framework that combines Gaussian Mixture Model (GMM)-based clustering with supervised learning to estimate the JND of each XR user from user-specific features. Leveraging the learned perception model, we formulate a perception-aware joint power and sub-band allocation problem and derive a Lyapunov-based solution that minimizes transmit power while satisfying probabilistic QoE constraints. Simulation results demonstrate that the proposed framework substantially improves radio resource efficiency compared with conventional QoS-driven schemes. In particular, the proposed method reduces the transmit power by up to approximately 26\% under stringent latency requirements and achieves power savings approaching 60\% in densely deployed in-body subnetworks, while maintaining the required QoE by allocating resources according to the actual perceptual capabilities of individual XR users rather than conservative application-level latency constraints.
\section{System Model and Problem Formulation}
\label{system-model}
\begin{figure}[t]
\centering
\includegraphics[width= 1\linewidth]{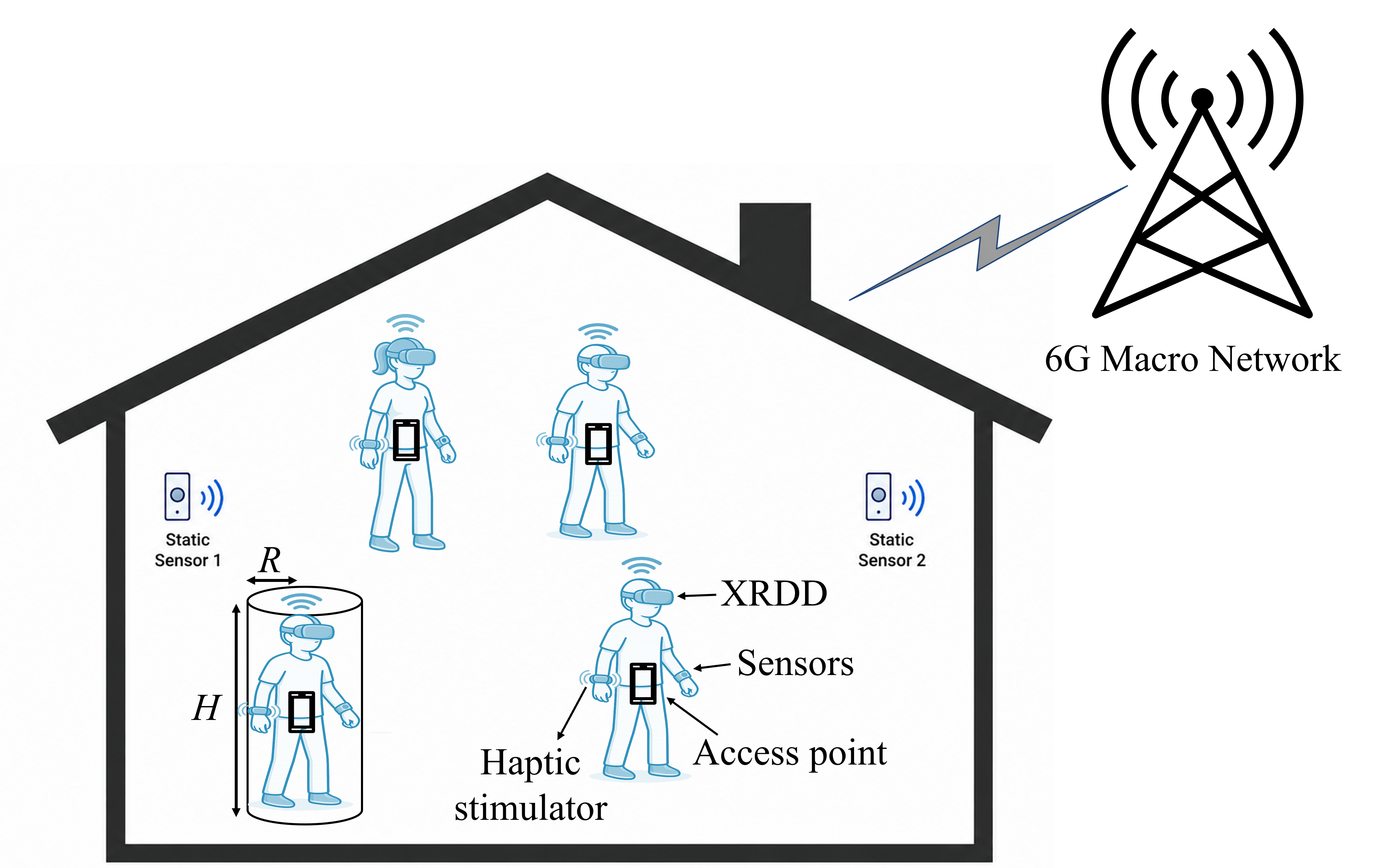}
\caption{Example of an indoor 6G subnetwork consisting of multiple XR users (IBSs).}
\label{network}
\end{figure}

We consider an indoor XR environment as depicted by Fig.~\ref{network} containing $N$ IBSs indexed by the set $\mathcal{N}=\{1, 2, ..., N\}$. Each IBS comprises an access point (AP), a set of wearable sensors, a haptic stimulator and an XR
display devices (XRDD). The AP collects sensory data (from sensors attached to the user's body), generates interactive XR video frames, and transmits them to the associated XRDD. In order to provide immersive XR experience, the edge server may also produce haptic feedback to the users, e.g., in the form of vibration, heating. We focus exclusively on the downlink transmission of high-quality XR video frames from the AP to the XRDDs. Specifically, we aim to minimize the average transmit power between the AP and XR users. The IBSs operate under the coverage of a macro 6G base station (BS) which handles the radio resource allocations of the APs.

As illustrated by Fig.~\ref{network}, each IBS is modeled as a cylindrical volume with radius $R$ and height $H$, within which the AP and XRDD are positioned at different heights. Both devices are equipped with a single antenna. The IBSs share a total bandwidth of $K$ sub-bands, represented by the set $\mathcal{K}$.  We consider a time-slotted system where each time slot $t$ has a fixed duration $\delta$. The channel between the $i$-th AP and the $n$-th XRDD at time slot $t$ and over the $j$-th sub-band is denoted by $h^j_{in}=\sqrt{\beta_{in}}g^j_{in}$, where $\sqrt{\beta_{in}}$ is the large-scale fading component including path loss and shadowing and $g^j_{in}$ is the small-scale Rayleigh fading component that follows the Jake's model~\cite{Jake} where the time correlation is given as $g^j_{in} (t+1)=\alpha g^j_{in}(t)+\sqrt{1-\alpha^2}\Delta^j_{in} (t+1)$, where $\alpha \in[0, 1]$ is the correlation coefficient of the channels over two consecutive time slots. Both $g^j_{in} (0)$ and $\Delta^j_{in} (t+1)$ follow a complex Gaussian distribution.

We consider the M/M/1 queuing model for the AP buffer. The video packets arrive at the AP buffer at each time slot according to an independent Poisson process with rate $a_i(t)$.  The packet size $L_i$ of the video packets follows an exponential distribution. Denoting $q_i(t)$ as the queuing delay at the $i$-th AP, the total delay (queuing and transmission) of user $i$ is $D_i(t)=q_i(t)+\frac{L_i}{r_i(t)}$, where $r_i(t)$ is the instantaneous achievable rate of the $i$-th AP which is given by
\begin{equation}\label{rate}
r_i(t)=B
\sum_{j\in\mathcal{K}}
\rho_{ij}(t)
\log_2
(
1+\eta_{ij}(t)),
\end{equation}
where $B$ is the sub-band bandwidth and $\eta_{ij}(t)$ is the signal-to-interference-plus-noise ratio (SINR) of the $i$-th IBS over the $j$-th sub-band at time $t$. $\rho_{ij}(t)$ is the sub-band allocation indicator where $\rho_{ij}(t)=1$ if sub-band $j$ is allocated to IBS $i$ at time slot $t$ and $\rho_{ij}(t)=0$, otherwise. Assuming that the sub-band bandwidth is considerably smaller than the coherence bandwidth, $\eta_{ij}(t)$ is given as
\begin{equation}
\eta_{ij}(t)
=
\frac{
p_{ij}^{AP}(t)\,
\rho_{ij}(t)
|
(
h_{ii}^j(t)
|^2
}{
I_{i,\mathrm{ibs}}^j(t)
+
\sigma_w^2
},
\label{eq:ibs_sinr}
\end{equation}
where $p_{ik}^{AP}(t)$ represents the transmit power of the $i$-th AP over sub-band $j$ subject to a maximum power of $P_{max}$ and $h_{ii}^j(t)$ is the channel between the AP and XRDD of the IBS $i$.  $\sigma_w^2$ denotes the additive white Gaussian noise (AWGN) power. The interference component $I_{i,\mathrm{ibs}}^j(t)$ captures the co-scheduling interference from other IBS users and is given by
\begin{equation}
I_{i,\mathrm{ibs}}^j(t)
=
\sum_{\substack{l=1\\l\neq i}}^{N}
p_{lj}^{AP}\,
\rho_{lj}(t)
|
h_{li}^j(t)
|^2.
\label{eq:ibs_interference}
\end{equation}

 We define $\tau_i(t)$ as the JND of IBS user $i$ at time slot $t$. Particularly, $\tau_i(t)$ defines perceptual sensitivity of user $i$ to the transmission delays of the XR videos. If the latency decreases below $\tau_i(t)$, the corresponding QoS improvement becomes imperceptible to the user and therefore provides negligible enhancement in the perceived QoE. In practice, $\tau_i(t)$ may vary over time depending on several factors, including user interaction intensity,  cognitive state, human fatigue and cognitive abilities and can be estimated using machine learning as will be explained in Section~\ref{JND}. Moreover, we define $D_i^{max}(\tau_i(t))$ as the maximum tolerable delay of user $i$, which depends on $\tau_i(t)$. Particularly, the IBS user is considered satisfied if  $\Pr\Big\{D_i(t)\geq
D_i^{\max}\big(\tau_i(t)\big)\Big\}\leq\epsilon_i\big(\tau_i(t)\big)$. The parameter $\epsilon_i(\tau_i(t))$ specifies the maximum  probability that the experienced delay exceeds $D_i^{max}(\tau_i(t))$. In this work, we develop a joint power and sub-band allocation algorithm with the aim of minimizing the average transmit power of the APs (IBSs)  considering the perception constraints (JND) of the XR users. By explicitly accounting for these human perception characteristics, the network can intelligently avoid unnecessary over-provisioning of radio resources and allocate bandwidth and power only when they result in meaningful perceptual gains.  The optimization problem is given as
\begin{subequations}
\label{opti}
\begin{align}
&\min_{\boldsymbol{\rho}(t), \boldsymbol{P}(t)}
\quad 
\sum_{j\in\mathcal{K}}
\sum_{i\in\mathcal{N}}
\bar{P}_i^{\,j}
,
\label{eq:2a}
\\
\textrm{s.t.}\quad
&
\Pr\Big\{
D_i(t)\geq
D_i^{\max}\big(\tau_i(t)\big)
\Big\}\leq
\epsilon_i\big(\tau_i(t)\big),\; \forall i\in\mathcal{N}
\label{eq:2b}
\\
&
0\leq p_{ij}^{AP}(t) \leq P_{max},\; \;
\forall i\in\mathcal{N},\; \forall j\in\mathcal{K}, 
\label{eq:2c}
\\
&
\rho_{ij}(t)\in\{0,1\}, \;  \forall i\in\mathcal{N}.
\; j\in\mathcal{K},
\label{eq:2d}
\end{align}
\label{eq:optimization_problem}
\end{subequations}
where $\boldsymbol{\rho}(t)$ is $N\times K$ matrix with each element $\rho_{ij}(t)$ and $\boldsymbol{P}(t)$ is $N\times K$ matrix having each element $p_{ij}^{AP}(t)$ as the instantaneous power allocated to IBS $i$ on sub-band $j$. The term $\bar{P}_i^{\,j}
=
\lim_{t\rightarrow\infty}
\frac{1}{t}
\sum_{x=0}^{t-1}
\rho_{ij}(x)p_{ij}(x)$ denotes the long-term average transmit power allocated to IBS user $i$ on sub-band $j$. Constraint~(\ref{eq:optimization_problem}b) ensures satisfying the required  QoE of each IBS user. Constraints~(\ref{eq:optimization_problem}c) and ~(\ref{eq:optimization_problem}d) enforce feasible power allocation and binary sub-band assignment decisions. Unlike standard resource allocation problems which enforce a predefined, application-centric latency constraint, our proposed method considers the human perceptual limitations $\left(\tau_i(t)\right)$ where the network can avoid wasting radio resources on improving QoS requirements (low transmission latency of video frames) that are not perceived by the user.
\section{Joint Power and Sub-band Allocation}\label{JND}
\subsection{Modeling and Estimation of the JND}
In order to solve the defined optimization problem in~\eqref{opti}, we first introduce a learning method to estimate $\tau_i(t)$ using a set of user features. The learning method combines both unsupervised learning (clustering) and supervised learning. We consider that each XR user has a time-varying feature vector $\mathbf{x}_i(t)$ comprising $d$ features (e.g., age, location, gaming/XR experience level). These features are assumed to be known to the 6G BS (e.g., collected via the different APs when the user registers in the network or by using sensors). We develop a  reliability-aware supervised learning algorithm that maps the features of the users to their JNDs, i.e., $\tau_i(t)=f(\mathbf{x}_i(t))$. Assume a dataset $\mathbf{x}_1(t),\ldots,\mathbf{x}_n(t)$ where $\mathbf{x}_i(t)\in\mathbb{R}^{d}$ contains $d$ user-related features. These features may include both numerical attributes, such as age, and categorical attributes, such as occupation or location. For each feature vector $\mathbf{x}_i(t)$ (input), there is an output value which is the corresponding JND $\tau_i(t)$. Such data can be collected through experiments or user studies designed to evaluate human delay sensitivity~\cite{dataset}. Since the temporal dependence can be handled separately using time-series analysis techniques~\cite{dataset2}, the time notation is omitted hereafter for simplicity. The dataset is represented by a matrix $\mathbf{X}\in\mathbb{R}^{n\times d},$ given as
\begin{equation}
\mathbf{W}
=
[\mathbf{X}~\boldsymbol{\tau}]
=
\begin{bmatrix}
\mathbf{w}_1^{T}
\\
\vdots
\\
\mathbf{w}_n^{T}
\end{bmatrix}
=
\begin{bmatrix}
\mathbf{x}_1^{T} & \tau_1
\\
\vdots & \vdots
\\
\mathbf{x}_n^{T} & \tau_n
\end{bmatrix},
\label{eq:W}
\end{equation}
where $\mathbf{w}_i\in\mathbb{R}^{d+1}$ is a vector contains both $\tau_i(t)$  and the associated user features. The first step is the unsupervised learning where  a Gaussian Mixture Model (GMM) is  fitted to the dataset using the Expectation-Maximization (EM) algorithm~\cite{EM}. Specifically, the EM algorithm is used to estimate the joint probability distribution $p(\mathbf{x},\tau_i(t))$, thereby enabling identification of the different perceptual modes (clusters) associated with the human brain. After estimating $p(\mathbf{x},\tau_i(t))$, each feature vector $\mathbf{x}_i$ is assigned a label $c_i\in \mathcal{C}=\{1,\ldots,m\}$, where $c_i$ represents a brain mode associated with $\tau_i(t)$. The identified clusters will serve as the labels for the dataset for the supervised learning phase. Specifically, the 6G BS will first classify the brain mode of the XR users joining the network, then the user's mode will be used to derive the probabilistic model of the associated  $\tau_i(t)$. As it has been shown that the delay perception of humans follows a multi-modal distribution~\cite{human1}, the distribution of the vector  $\mathbf{w}_i$ can be given as~\cite{vector}
\begin{equation}
p(\mathbf{w})
=\sum_z p(z) p(\mathbf{w}|z)=
\sum_{k=1}^{m}
\pi_k
f(\mathbf{w}|\boldsymbol{\mu}_k,\boldsymbol{\Sigma}_k),
\label{eq:gmm}
\end{equation}
where $f(\mathbf{w}|\boldsymbol{\mu}_k,\boldsymbol{\Sigma}_k)$ denotes the multivariate Gaussian probability density function with mean vector $\boldsymbol{\mu}_k$ and covariance matrix $\boldsymbol{\Sigma}_k$. $z$ is a binary random vector where a particular element $z_k=1$ and all other elements equal to 0. The parameter $\pi_k=p(z_k=1)=$ represents the mixing coefficient associated with mode $k$. Accordingly, the delay perception behavior of a human user is modeled as belonging to mode $k$ with probability $\pi_k$. The posterior probability, also referred to as the responsibility of mode $k$, is given by
\begin{equation}
r_i(z_k)
=
\frac{
\pi_k
f(\mathbf{w}|\boldsymbol{\mu}_k,\boldsymbol{\Sigma}_k)
}{
\sum_{j=1}^{m}
\pi_jf(\mathbf{w}|\boldsymbol{\mu}_j,\boldsymbol{\Sigma}_j)
}.
\label{eq:responsibility}
\end{equation}
Each data point is assigned to the mode having the highest posterior probability. The EM algorithm iteratively estimates the GMM parameters ($\boldsymbol{\mu}_k$ and $\boldsymbol{\Sigma}_k$) using the real-time cognitive behavior of the user~\cite{EM}. The corresponding log-likelihood function is expressed as
\begin{equation}
\ln L
(
\boldsymbol{\Sigma},
\boldsymbol{\mu},
\boldsymbol{\pi}
|
\mathbf{w}
)
=
\sum_i
\ln
\sum_{k=1}^{m}
\pi_k
f(\mathbf{w}_i|\boldsymbol{\mu}_k,\boldsymbol{\Sigma}_k).
\label{eq:likelihood}
\end{equation}
Since the likelihood function in~\eqref{eq:likelihood} has singularities,
it is infeasible to find $\pi_k$, $\boldsymbol{\mu}_k$ and $\boldsymbol{\Sigma}_k$. Therefore, we utilize the EM algorithm proposed in~\cite{EM2} that maximizes the likelihood function for a GMM. The EM procedure initializes the parameters randomly and subsequently alternates between evaluating the responsibilities (using \eqref{eq:responsibility}) and updating the GMM parameters to maximize the likelihood function in~\eqref{eq:likelihood}. Based on the GMM, the dataset is clustered (labeled).  For each feature vector $\mathbf{w}_i$, the corresponding label is determined according to
\begin{equation}
c(\mathbf{w}_i)
=
\arg\max_k
p(z_k=1|\mathbf{w}_i).
\label{eq:cluster}
\end{equation}
%
The resulting labels are then used to train a supervised learning classifier that maps the user feature vectors to the corresponding perceptual modes. Denoting $\mathbf{y}=[c(\mathbf{w}_1), \cdots,c(\mathbf{w}_n)]^T$ as the output vector of the GMM, the supervised learning builds the model $f$ that approximates the mapping $c_i=f(\mathbf{x}_i)$, which identifies the perceptual mode associated with a given feature vector. Once the classifier is trained, the perceptual mode of each user can be identified directly from its observed features. Based on the identified mode, a probabilistic model for the JND can then be derived. To characterize this relationship, we define the effective delay $D_i^{\min}(\epsilon')$ for user $i$ where
\begin{equation}
\Pr
\left\{
\tau_i(t)
<
D_i^{\min}(\epsilon')
\right\}
<
\epsilon'.
\label{eq:effective_delay}
\end{equation}
The effective delay represents the minimum delay value that remains imperceptible to the user with probability $1-\epsilon'$. The following Theorem~1 defines the relation between $D_i^{\min}(\epsilon')$ and $\epsilon'$. For simplicity, we use $D_i^{\min}$ hereafter.

\textbf{Theorem 1:} Assuming that mode $k$ has been identified for user $i$, $\tau_i(t)$ can be bounded according to
\begin{equation}
\Pr
\left\{
\tau_i(t)
<
\mu_k({d+1})
-
\sqrt{
Q_{d+1}(\gamma)
\mathbf{e}_{d+1}^{T}
\boldsymbol{\Sigma}_k
\mathbf{e}_{d+1}
}
\right\}
<
\frac{1-\gamma}{2},
\label{eq:bound}
\end{equation}
where $\boldsymbol{\Sigma}_k$ and $\mu_k({d+1})$ represent the covariance matrix and  the ($d+1$)th element of mean vector of the brain mode $k$, respectively. $\mathbf{e}_j$ is a unit vector in $\mathbb{R}^{d+1}$ whose $j$th element is equal to $1$ while all remaining elements are equal to $0$. $Q(\gamma)$ represents the quantile function of the chi-square distribution which is given by
\begin{equation}
    Q_{d+1}(\gamma) = \inf \left\{
x \in \mathbb{R}
\;\middle|\;
\gamma \leq
\int_{0}^{x}
\chi^{2}_{d+1}(r)\,dr
\right\},
\end{equation}
where $d$ denotes the number of features used for learning, and $\chi^2_{d+1}(x)$ is the probability density function of a chi-square random variable with $(d+1)$ degrees of freedom. Then, $D_i^{\min}$ can be given as
\begin{equation}
D_i^{\min}
=
\mu_k(d+1)
-
\sqrt{
Q_{d+1}(1-2\epsilon')
\mathbf{e}_{d+1}^{T}
\Sigma_k
\mathbf{e}_{d+1}
},
\end{equation}

\textit{Proof:} The proof is omitted due to space limitation.

Based on the fact that the reliability of the system is $1-(\epsilon+\epsilon')$, we can set the values of $\epsilon$ and $\epsilon'$, then, we can find the tolerable perception delay $D_i^{\max}(\tau_i(t))$  as  $D_i^{\max}(\tau_i(t))=D_i^{\min}(\epsilon')$.

\subsection{Optimal Resource allocation via Lyapunov optimization}
\label{proposed}
With the perception model specified, the optimization problem in~(\ref{eq:optimization_problem}) becomes fully specified. Nevertheless, solving the formulated problem remains highly challenging due to the stochastic and time-varying nature of both the wireless channel conditions and the users’ JND $\tau_i(t)$. In particular, the probabilistic delay constraint in~(\ref{eq:2b}) becomes difficult to handle directly because the achievable data rates and the perception-aware delay requirements evolve dynamically over time. Therefore, a tractable reformulation of the original constraint is required before deriving an efficient radio resource allocation strategy. The objective is to characterize the probability that the experienced delay exceeds the maximum tolerable delay threshold $D_i^{\max}$. Without loss of generality,we assume that packet arrivals follow a Poisson process with the rate $a_i(\nu)$ and the user's data rate is exponentially distributed with a parameter $r_i(\nu)$ at time slot $\nu=1, \cdots, t$. We define the packet service time as the transmission time of the packet from the AP to the XRDD. Consider that the packet lengths follow an exponential distribution with parameter  $\chi$. If the IBS is allocated a fixed rate $r_i$, then, the service time will also follow an exponential distribution with parameter $\chi r_i$, i.e., the PDF of the service time $s$ is $e^{-\chi r_is}$. Without loss of generality, and considering constant $\chi$, we assume that the service time of each packet is an exponential random variable with parameter $r_i$.

\textbf{Theorem 2:}
Consider a given user $i$ with a time-varying service  $r_i(\nu)$ at time slot $\nu$. If the duration of each time slot is sufficiently long such that the queue reaches steady state, i.e.,
\begin{equation}
\frac{1}{r_i(\nu)-a_i(\nu)}
\ll
\delta\nu,
\label{eq:11}
\end{equation}
then the probability that the experienced delay exceeds the threshold $D_i^{\max}$ is given by
\begin{equation}
\Pr(D_i>D_i^{\max})
=
\lim_{t\rightarrow\infty}
\frac{1}{t}
\sum_{\tau=1}^{t}
e^{-\left(r_i(\nu)-a_i(\nu)\right)D_i^{\max}},
\label{eq:12}
\end{equation}
under the condition that $r_i(\nu)>a_i(\nu), \quad \forall \nu>0$.

\textit{Proof:} The proof is omitted due to space limitation.

Theorem~1 implies that the probabilistic delay constraint in~(\ref{eq:2b}) can be satisfied whenever the following condition holds:
\begin{equation}
\lim_{t\rightarrow\infty}
\frac{1}{t}
\sum_{\nu=1}^{t}
e^{-\left(r_i(\nu)-a_i(\nu)\right)D_i^{\max}}
<
\epsilon,
\label{eq:13}
\end{equation}
which transforms the original probabilistic latency requirement into a more tractable time-average form. The reformulated constraint in~(\ref{eq:13}) naturally aligns with the Lyapunov drift-plus-penalty optimization framework. Since both the wireless channel gains $h_{ij}(t)$ and JND $\tau_i(t)$ evolve dynamically over time, the optimization problem must be solved repeatedly at each time slot. During each slot, the channel gains and JNDs are assumed to remain fixed.

The drift-plus-penalty framework enables stabilization of queue dynamics while simultaneously minimizing a long-term network cost function. To guarantee satisfaction of the perception-aware latency constraint, a virtual queue is introduced for each user as
\begin{equation}
F_i(t+1)
=
\max
\left\{
F_i(t)
+
e^{-\left(r_i(t+1)-a_i(t+1)\right)D_i^{\max}}
-
\epsilon, 0
\right\}.
\label{eq:14}
\end{equation}
If the virtual queue $F_i(t)$ is mean-rate stable, namely,
\[
\lim_{t\rightarrow\infty}
\frac{F_i(t)}{t}
=
0,
\]
then the following condition is satisfied:
\begin{equation}
\lim_{t\rightarrow\infty}
\frac{1}{t}
\sum_{\tau=1}^{t}
e^{-\left(r_i(\nu)-a_i(\nu)\right)D_i^{\max}}
<
\epsilon.
\label{eq:15}
\end{equation}

To jointly guarantee queue stability and minimize the transmit power consumption, the Lyapunov function associated with the virtual queues is defined as
\begin{equation}
L(t)
=
\frac{1}{2}
\sum_{i\in\mathcal{N}}
F_i^2(t).
\label{eq:16}
\end{equation}

Subsequently, the total transmit power of the BS is adopted as the penalty function. By minimizing the drift-plus-penalty expression, the original optimization problem is transformed into the following per-slot optimization problem:
\begin{subequations}
\begin{align}
&\min_{\boldsymbol{\rho}(t), \boldsymbol{P}(t)}
\;
V
\sum_{i,k}
p_{ij}^{AP}(t)
+
\sum_{i\in\mathcal{N}}
y_i(t)F_i(t),
\label{eq:17a}
\\
\textrm{s.t.}\quad
&
r_i(t)>a_i(t),
\label{eq:17b}
\\
&
0\leq p_{ij}^{AP}(t) \leq P_{max},\; \;
\forall i\in\mathcal{N},\; \forall j\in\mathcal{K},
\label{eq:17c}
\\
&
\rho_{ij}(t)\in\{0,1\},
\;
\forall i\in\mathcal{N},
\;
j\in\mathcal{K},
\label{eq:17d}
\end{align}
\label{eq:17}
\end{subequations}
where
\[
y_i(t)
=
e^{-\left(r_i(t)-a_i(t)\right)D_i^{\max}}
-
\epsilon.
\]

The objective function in~(\ref{eq:17a}) jointly captures the transmit power minimization objective and the perception-aware delay reliability requirement and is equivalent to \ref{eq:optimization_problem}a) and \ref{eq:optimization_problem}b) in the original problem. Constraint~(\ref{eq:17b}) guarantees queue stability i.e., the queue length does not approach
infinity which is essential to satisfy constraint \ref{eq:optimization_problem}b). The other constraints remain the same as those in \ref{eq:optimization_problem}). Since the resulting optimization problem remains coupled across sub-bands, we employ decomposition techniques and low-complexity approximation methods (e.g., C-additive approximation) and solve the problem using the ellipsoid method~\cite{decomp}.

\section{Performance Evaluation}
\label{results}
We consider a network with total bandwidth of 100 MHz divided into $K=4$ sub-bands. The performance of the proposed method is assessed through a system-level simulator developed in accordance with the 3GPP evaluation methodology~\cite{simulation}. The locations of the IBSs are generated according to a Thomas Cluster Process (TCP), comprising five cluster centers and a standard deviation of $2$~m for the displacement of offspring points from their associated cluster centers~\cite{TCP}. Each IBS is represented by a cylindrical volume with a radius of $0.25$~m and a height of $1.9$~m, where the cylinder center denotes the IBS position. To prevent excessive overlap among deployments, a minimum distance of $4$~m is maintained between cluster centers, while a minimum separation of $0.5$~m is imposed between IBS centers.  The main simulation parameters are provided in Table~\ref{t1}. Distinct channel models are employed to characterize intra-IBS and inter-IBS propagation conditions. For intra-IBS links, the large-scale fading component is modeled as $\beta_{nn}(\mathrm{dB}) = 8.6\log_{10}(r_{nn}) + 46.1 + 2\chi_{nn}$, where \(\chi_{nn} \sim \mathcal{N}(0,1)\) denotes the log-normal shadowing term~\cite{Takizawa2008WBAN} and \(r_{nn}\) represents the distance between the AP and XRDD of the \(n\)-th IBS. For inter-IBS communications, we adopt the 3GPP D2D propagation model~\cite{3GPP36843}. The large-scale fading coefficient is expressed as $\beta_{nl}(\mathrm{dB}) = 38.8 + 16.9\log_{10}(r_{nl}) + 3\chi_{nl}$, for line-of-sight (LOS) links, and as $\beta_{nl}(\mathrm{dB}) = 17.5 + 43.3\log_{10}(r_{nl}) + 4\chi_{nl}$ for non-line-of-sight (NLOS) links. Here, \(r_{nl}\) denotes the three-dimensional separation between the \(n\)-th XRDD and the \(l\)-th AP. The LOS probability is modeled as $\Pr\{S_{nl}=\mathrm{LOS}\}=\exp(-\Gamma r'_{nl})$, where \(\Gamma = 2\lambda R\), $\lambda$ is the IBS density, and \(r'_{nl}\) corresponds to the two-dimensional distance between the \(l\)-th AP and the \(n\)-th XRDD. This formulation is derived by treating the IBS cylinders as potential blockages affecting inter-body wireless links. 
\begin{table}[t!]
		\centering
		\caption{Simulation parameters}
		\label{t1}
		\begin{tabular}{ll}
			\toprule
			Parameter & Value \\
				\midrule
			Total bandwidth & 100 MHz\\
            Number of sub-bands & 4\\
            Maximum power (dBm) & 10\\
            IBS cylinder dimensions (m) & $R=0.25$, $H=1.9$ \\
            $\sigma_w^2$  & -180 dBm\\
            $a_i(t)$  & 45 Mbps\\
            $\epsilon$  & 0.05\\
			\bottomrule
		\end{tabular}	
	\end{table}

To characterize the delay perception of IBS users, we use the dataset in~\cite{dataset}. In \cite{dataset}, the authors conducted a user study involving 30 participants. Each participant was asked to evaluate the quality of five video sequences while the system delay and packet loss levels were progressively increased. We use the average quality ratings provided by each participant to estimate their individual JND. Specifically, the delay perception of a user is defined in~\cite{dataset} as the maximum delay that the user is unable to perceive, which is consistent with the delay perception definition adopted in this work. Since no publicly available dataset simultaneously captures per-user feature vectors alongside subjective delay perception profiles, we synthetically construct three continuous features per user as follows. First, the delay perception functions $\tau_i(t)$ are clustered across a population of 1000 users. For each resulting cluster, a random mean vector and a random positive semidefinite covariance matrix are sampled and used to generate multivariate Gaussian feature realizations for every user within that cluster. By construction, the synthesized features exhibit two key properties: (1) a GMM marginal distribution over the entire user population, and (2) statistical correlation with, and hence predictive relevance for, the individual delay perception $\tau_i(t)$.
\begin{figure}[t]
\centering
\includegraphics[width= 0.8\linewidth]{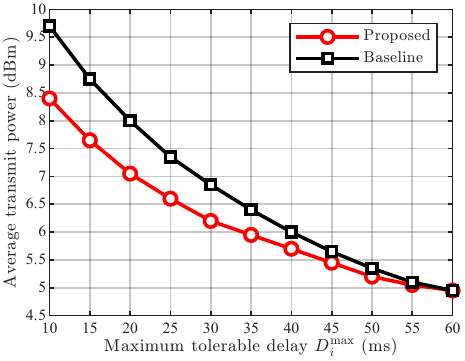}
\caption{Average transmit power of an AP versus maximum tolerable delay $D_i^{\max}$ with $N=10$.}
\label{fig:avg_power_delay}
\end{figure}

Figure~\ref{fig:avg_power_delay} illustrates the average transmit power per user (i.e., AP) for $N=10$ (randomly selected from the dataset) as a function of the maximum tolerable delay $D_i^{\max}$ for our proposed scheme and a baseline. The adopted baseline is a typical delay-aware power allocation  scheme where the power is allocated based on a fixed delay constraint (i.e., constraint~\eqref{eq:2b} is fixed). In the baseline method, both $D_i^{\max}$ and $\epsilon$ are independent of $\tau_i(t)$. As shown by Fig.~\ref{fig:avg_power_delay}, the average transmit power decreases monotonically with increasing $D_i^{\max}$ for both schemes, since a more relaxed delay constraint \eqref{eq:2b} is easier to satisfy and consequently requires less transmit power. As $D_i^{\max}$ increases to 60~ms, the two schemes converge and the power savings from our proposed approach become negligible as $D_i^{\max}$ and $\tau_i(t)$ become close to each other and the perceptual constraint no longer offers an exploitable slack. However, the advantage of our proposed approach is most pronounced under stringent latency regimes, which is the most resource-intensive and practically relevant operating regime for emerging XR applications. For instance, at $D_i^{\max}=10$~ms, the proposed scheme reduces the transmit power from 9.75~dBm to 8.44~dBm, equivalent to approximately 26\% less power compared to the baseline. This is mainly due to the ability of the proposed scheme  to exploit the learned JND $\tau_i(t)$, to avoid allocating unnecessary resources to meet a QoS target that is imperceptible to the user.
\begin{figure}[t]
\centering
\includegraphics[width= 0.8\linewidth]{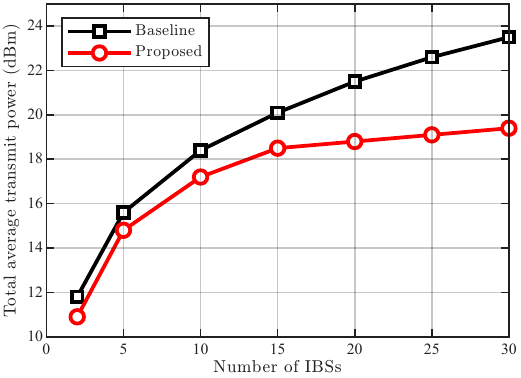}
\caption{Total average transmit power versus the number of IBSs ($N$) with $D_i^{\max}=20$ ms.}
\label{fig:avg_totalpower}
\end{figure}

Fig.~\ref{fig:avg_totalpower} evaluates the performance of the two schemes as a function the network size by showing the total average transmit power in~\eqref{eq:2a} as a function of the total number of IBSs. Both curves rise with network load, reflecting the well-known bandwidth-per-user trade-off: as more users compete for the same spectral resources, the system must expend more power to maintain the delay requirement of each user. Notably, the perception-aware scheme outperforms its counterpart across the entire load range, and the margin widens substantially as the network becomes congested. With $N=2$, the difference is modest (approximately 1 dBm), whereas at $N=30$ it grows to roughly 4 dBm, equivalent to a power saving of approximately 60\%. This scaling behavior arises because, under high network load, the ability to tailor resource allocation to each user's actual JND $\tau_i(t)$ rather than  provisioning resources against a conservative, application-level QoS target for every user. 
\begin{figure}[t]
\centering
\includegraphics[width= 0.8\linewidth]{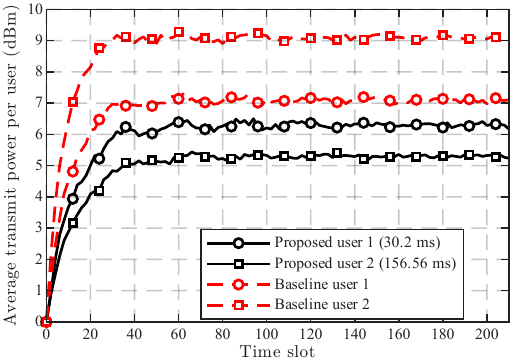}
\caption{Convergence of the average transmit power for two representative IBS users under the proposed and baseline schemes.}
\label{fig:learning}
\end{figure}

In Fig.~\ref{fig:learning}, we show the evolution of the average transmit power per user over time. In this figure, the power allocations of two IBSs (randomly selected) follow the proposed perception-aware method (i.e., two users are selected to learn their JNDs. The other 8 IBSs follow the baseline method with predefined $D_i^{\max}$. The learned JNDs ($\tau_i(t)$) of the two users are 30.2~ms and 156.56~ms, respectively. The results given by Fig.~\ref{fig:learning} reflect two points. First, it shows that the proposed method successfully allocates power according to the distinct JNDs of the users, as we can see that  a user with a higher JND ($\tau_i(t)=156.56$~ms) will be allocated less power than a user with a lower JND ($\tau_i(t)=30.2$~ms). Second, the power consumption of the other users with predefined delay requirements is higher and differs mainly due to the physical channel gains, ignoring human factors and resulting in inefficient resource allocation.

\section{Conclusion}
\label{sec:conclusions}
This paper proposed a human-aware power control framework for XR-oriented 6G in-body subnetworks by incorporating users' delay perception into radio resource allocation. A machine learning framework was developed to estimate the perceptual delay threshold of XR users from user-specific features, and the learned model was integrated into a perception-aware joint power and sub-band allocation problem solved using a Lyapunov drift-plus-penalty framework. Simulation results demonstrated that the proposed approach improves radio resource efficiency by significantly saving the average transmit power while maintaining acceptable QoE in  challenging, dense deployment scenarios. This work can be extended by conducting experimental validation using real XR users and measured delay perception datasets to further assess the practical benefits of perception-aware radio resource management.
	\bibliographystyle{IEEEtran}
\bibliography{main}
	
\end{document}